\documentclass[ 
aps, 
prb,
reprint, 
twocolumn, 
superscriptaddress, 
showpacs,
hidelinks]{revtex4-1}

\usepackage{amsmath}
\usepackage{amssymb}
\usepackage{amsfonts}
\usepackage{dsfont}
\usepackage{bm}
\usepackage{mathrsfs}
\usepackage{mathtools}
\usepackage{graphicx}
\usepackage{hyperref}
\usepackage{xcolor}

\usepackage[english]{babel}
\usepackage[T1]{fontenc}
\usepackage[utf8]{inputenc}
\usepackage{color}

\newcommand{\up}{\uparrow}
\newcommand{\down}{\downarrow}

\newcommand{\calF}{\mathcal{F}}

\newcommand{\calJ}{\mathcal{J}}
\newcommand{\calL}{\mathcal{L}}
\newcommand{\calO}{\mathcal{O}}
\newcommand{\calP}{\mathcal{P}}
\newcommand{\calQ}{\mathcal{Q}}
\newcommand{\calR}{\mathcal{R}}

\newcommand{\calU}{\mathcal{U}}
\newcommand{\idop}{\mathds{1}}

\newcommand{\dd}[1]{\mathrm{d}#1\,}
\newcommand{\tot}{\mathrm{tot}}
\newcommand{\HLL}{\mathrm{HLL}}
\newcommand{\vF}{v_F}
\newcommand{\kbT}{k_B T}
\DeclareMathOperator{\Tr}{Tr}

\definecolor{BP-color}{named}{green}
\definecolor{PR-color}{named}{magenta}
\definecolor{PV-color}{rgb}{0.97,0.57,0.11}

\begin{document}
\title{Sub- to Super-Poissonian crossover of current noise in helical edge
  states coupled to a spin impurity in a magnetic field}

\author{Benedikt Probst}

\affiliation{Institut f\"ur Mathematische Physik, Technische Universit\"at
  Braunschweig, D-38106 Braunschweig, Germany}

\author{Pauli Virtanen}
\affiliation{Department of Physics and Nanoscience Center, University of Jyv\"askyl\"a,
  P.O. Box 35 (YFL), FI-40014 University of Jyv\"askyl\"a, Finland}

\author{Patrik Recher} 
\affiliation{Institut f\"ur Mathematische Physik, Technische Universit\"at
  Braunschweig, D-38106 Braunschweig, Germany} 
\affiliation{Laboratory for Emerging Nanometrology Braunschweig, D-38106
  Braunschweig, Germany}

\begin{abstract}
  Edge states of two-dimensional topological insulators are helical and
  single-particle backscattering is prohibited by time-reversal symmetry. In
  this work, we show that an isotropic exchange coupling of helical edge states (HES) to a
  spin 1/2 impurity subjected to a magnetic field results in characteristic
  backscattering current noise (BCN) as a function of bias voltage and tilt
  angle between the direction of the magnetic field and the quantization axis of
  the HES. In particular, we find transitions from sub-Poissonian (antibunching)
  to super-Poissonian (bunching) behavior as a direct consequence of the
  helicity of the edge state electrons. We use the method of full counting
  statistics within a master equation approach treating the exchange coupling
  between the spin-1/2 impurity and the HES perturbatively. We express the BCN
  via coincidence correlation functions of scattering processes between the HES
  which gives a precise interpretation of the Fano factor in terms of bunching
  and antibunching behavior of electron jump events. We also investigate the
  effect of electron-electron interactions in the HES in terms of the
  Tomonaga-Luttinger liquid theory.
\end{abstract}

\date{\today}

\maketitle
\section{Introduction}
\label{sec:Introduction}

\begin{figure}
  \centering
  \includegraphics{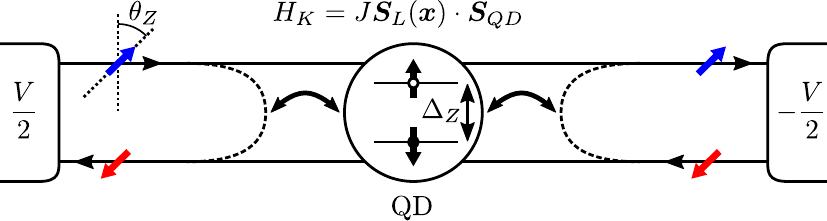}
  \caption{
    Sketch of the setup: An HLL is coupled to a QD containing a spin 1/2. The coupling between the HLL and the QD is governed
    by an isotropic Kondo Hamiltonian $H_K$. A magnetic field ${\bm B}$ is applied to the QD,
    which is tilted with respect to the quantization axis $\hat z$ of the helical
    edge state by an angle $\theta_Z$.
  }
 \label{fig:setup}
\end{figure}

Helical edge states (HES) are one of the hallmarks of the quantum spin Hall insulator (QSHI) realized in two-dimensional topological insulators (TIs) \cite{Kane2005, Bernevig2006, Liu2008}. The motion of charge and spin is locked and time-reversal symmetry (TRS) protects the electron flow from backscattering. Transport along the helical edge is therefore ballistic leading to a conductance of $e^2/h$ per edge in a two-terminal experiment \cite{Koenig2007, Roth2009, Knez2011}. Unlike in the quantum Hall effect where different transport directions are spatially separated by the insulating bulk, in the QSHI left- and right-movers exist on the same edge. Breaking TRS \cite{Xu2006, Wu2006, Piatrusha2019} or allowing for electron-electron interactions \cite{Wu2006, Strom2010, Schmidt2012, Crepin2012, Novelli2019}, electron-phonon interactions \cite{Budich2012, Groenendijk2018} or interactions with nuclei \cite{DelMaestro2013} as well as with nearby charge puddles \cite{Vayrynen2013, Aseev2016, Nagaev2018} induces backscattering corrections to the ballistic conductance value. Understanding such mechanism can explain some experimental deviations from the ballistic value but can also give insight to the nature of the helical edges states. Another prominent example involving conductance corrections to the ballistic value is the Kondo effect of a spin 1/2 impurity exchanged coupled to the HES \cite{Maciejko2009, Tanaka2011, Posske2013, Altshuler2013}.
We have previously analyzed the backscattering conductance from a spin 1/2 impurity subject to a magnetic field and weakly coupled to HES as a function of bias voltage and tilt-angle $\theta_Z$ of the magnetic field with respect to the HES' spin quantization axis (see Fig.~\ref{fig:setup}) \cite{Probst2015}. We found some characteristic resonance behavior in the conductance as a function of bias voltage $V$ when $eV$ becomes comparable to the Zeeman field. The conductance is strongly asymmetric under reversal of the bias voltage, the degree depending on the tilt angle. This is a strong sign of the helicity of the edge states.

The dynamics of transport processes can be better understood by looking at the current noise which is sensitive to correlations between backscattering events. \cite{Blanter2000} Bunching effects due to correlations between cotunneling processes are well studied in quantum dots and molecular magnets \cite{Sukhorukov2001,Thielmann2005}. Noise from backscattering events off a spin impurity in helical liquids has been considered so far in the absence of magnetic fields, where the Fano factor is always larger than one, pointing at bunching behavior \cite{Vayrynen2017, Nagaev2018, Kurilovich2019, Pashinsky2020}. Here, we show that in the presence of a Zeeman-field $\Delta_Z$ on the quantum dot, the bunching behavior becomes supplemented by an anti-bunching behavior for small tilt angles of the magnetic field for bias voltages that satisfy $k_BT < \Delta_Z\lesssim eV$ (for only one polarity of the bias voltage) where $T$ is the temperature. At larger voltages $|eV|\gg \Delta_Z$, we find super-Poissonian behavior for all tilt-angles.

The main results and its relation to the helical nature of the edge states can be understood as follows. Consider a general tilt angle $0<\theta_Z < \pi/2$. At bias voltages $|eV|< \Delta_Z$, the spin impurity will stay in the ground state $|\downarrow\rangle$ and backscattering events which transfer electrons between right- and left-moving channels are elastic and therefore uncorrelated which leads to Poissonian noise (Fano factor $F=1$). More interesting is the case $|eV|\gtrsim\Delta_Z$, where the impurity spin can get excited by backscattering events. Due to the helical nature of the edge states, the rate to flip the QD spin with the associated change of the spin in the helical leads also crucially depends on the spin bias in the helical lead that is a topological feature of the bulk of the TI. As a consequence, for one polarity of the bias voltage, the rate for relaxation of the QD spin is always larger than the rate for excitation. On the contrary, for the other polarity, there is a bias voltage where the rate for excitation starts to dominate the rate for relaxations. As a result, for large absolute values of the bias voltage, a slow spin flip process is followed by a fast spin flip process which leads to super-Poissonian noise and a Fano factor $F>1$. For $eV \gtrsim\Delta_Z$, the rate for excitations out of the spin ground state  $|\downarrow\rangle$ becomes finite and starts to polarize the QD spin in the excited state. We show by detailed investigations of conditioned scattering events that there is a region of bias voltages where the dominant scattering event is a fast excitation process followed by a slow relaxation process leading to sub-Poissonian noise with $F<1$. However, with increasing bias voltage the spin becomes polarized on average predominantly in the excited state $|\uparrow\rangle$ blocking the fast excitation rate and the dominant scattering process becomes the slower relaxation followed by an excitation with faster rate signalling the crossover to the super-Poisonian noise regime ($F>1$) at larger bias. This crossover noise behavior is most pronounced for small tilt angles $\theta_Z$ and disappears for $\theta_Z = \pi/2$ where the noise becomes symmetric as a function of bias voltage.

The rest of the article is structured as follows: In Section II, we introduce a general scheme based on full counting statistics and master equations
to formulate the average current and noise at zero frequency of electrons in a reservoir (bath) coupled to a system with a few degrees of freedom. We show in particular how the noise can be formulated in terms of coincidence functions (equal-time Glauber coherence functions) for general scattering events transporting any integer $n$ of electron charges at general bias voltage configurations thereby generalizing earlier results. In Section III, we consider the explicit example of the spin 1/2 impurity (system) subjected to a Zeeman field in a general direction with respect to the spin quantization axis of the helical edge states (the bath). We derive expressions for current and noise in terms of steady state quantum statistical averages of jump operators, discuss in detail the backscattering rates and explain the characteristic bunching and antibunching behavior in terms of coincidence functions and with the density matrix conditioned on certain scattering events. We also comment on the effect of Luttinger liquid correlations in interacting helical liquids.

\section{Transport and Statistics}
\label{sec:Model}

\subsection{Master equation approach to Full Counting Statistics}

The correlations of scattering processes of electrons between the
helical edge states in the proposed setup can be suitably described by
Glauber correlation functions known from quantum optics,
\cite{Glauber1963,Glauber2006} which in the electronic context have been
previously used e.g. in discussions of single-electron sources \cite{Feve2007,Haack2013} and can
be employed to characterize the frequency-dependent noise and bunching in transport.
\cite{Saito1992,Blanter2000,Emary2012}

\begin{figure}
  \centering
  \includegraphics[width=5cm]{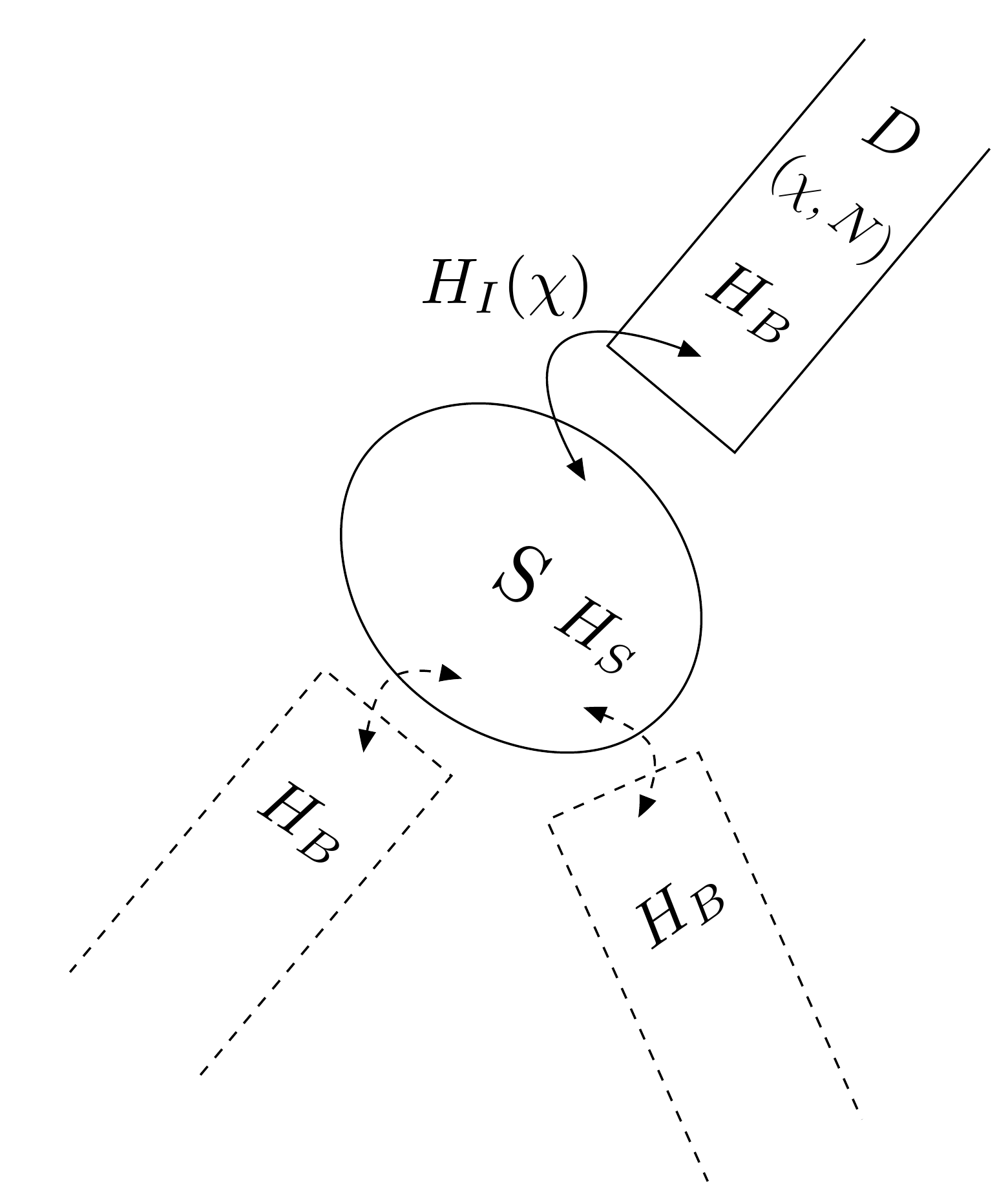}
  \caption{ Sketch of the counting setup where leads (bath $B$ with Hamiltonian $H_B$) is coupled to a system ($S$ with Hamiltonian $H_S$) by a coupling Hamiltonian $H_I(\chi)$ that depends on the counting field $\chi$ that is the conjugate variable to the particle number $N$ of the drain ($D$). The dashed lines denote possible additional leads and couplings.}
 \label{fig:setup_master}
\end{figure}

In this section, we apply the method of electron full counting statistics (FCS)
\cite{Levitov1996,*Nazarov2003,*Levitov2004}
to derive precise formulas for the current noise in terms of delayed coincidence correlation functions (equal-time Glauber functions).
For completeness and to introduce notation, we recall how to introduce
a cumulant generating function
(CGF) and derive it from a master equation containing the counting
field considering a quite general system comprised of particle reservoirs (the bath), a system and a coupling between the system and the reservoirs (see Fig.~\ref{fig:setup_master}).
From the long time behavior of the reduced density matrix for the system, we will then calculate the average current and the zero-frequency current noise in one of the reservoirs (the drain) perturbatively in the counting field. 
In general this formalism can be extended to cover multi-terminal setups.\cite{Bagrets2003}

The CGF \cite{Levitov1996,*Nazarov2003,*Levitov2004}  $S(\chi,t)$ is defined via a counting field $\chi$ 
 \begin{equation}
 \label{CGF1}
 e^{-S(\chi,t)}=\sum_n P(n,t) e^{i n \chi},
\end{equation}
where $P(n,t)$ is the probability that $n$ particles have been added to the drain during a time interval $t$. Quantum mechanically, the CGF can also be written as
a trace \cite{NoiseNatoBook2003}
\begin{equation}
\label{CGF2}
 e^{-S(\chi,t)}=\Tr\Big(\rho_\tot(\chi;t)\Big),
\end{equation}
where the dynamics of the density matrix of the
total system is
\begin{equation}
 \label{eq:rhoDynamicsWithCountingField}
 \rho_\tot(\chi;t)=\calU\big(\chi/2;t,0\big)\rho_{\tot,0}\calU\big(-\chi/2;0,t\big),
\end{equation}
with $\rho_{\tot,0}$ being the initial density matrix and the propagator is
defined by
\begin{equation}
  \label{eq:DefPropWithCF}
  \calU(\chi;t,t')
  =
  e^{i\chi\hat N}\calU(t,t')e^{-i\chi\hat N},
\end{equation}
where $\hat N$ is the number operator for the electrons in the drain and
$\calU(t,t')$ is the time propagator. Using the representation of $\calU(t,t')$
by a time ordered exponential we find
\begin{equation}
  \label{eq:DefPropAsTimeOrderedExp}
  \calU(\chi;t,t')=
    {\hat T}\exp\bigg(-\frac{i}{\hbar}\int_{t'}^{t}H(\chi)\dd\tau\bigg),
\end{equation}
where ${\hat T}$ is the time ordering operator and we used that the counting field does not
depend on time such that the counting field dependence can be included in the
Hamiltonian
\begin{equation}
 H(\chi)=e^{i\chi\hat N}He^{-i\chi\hat N},
\end{equation}
where $H$ is the Hamiltonian of the total system. Note that according to Eq.~(\ref{CGF2}), we only need the reduced density matrix $\rho(\chi;t)$ of the system defined by tracing over the bath $\rho(\chi;t)={\rm Tr}_{\rm B} (\rho_\tot(\chi;t))$, so that  $\exp(-S(\chi,t))={\rm Tr}(\rho(\chi;t))$, where the trace now is only taken in the system subspace. Using the Fourier transformation $\rho(\chi;t)=\sum_n \rho_n(t)\exp(i n\chi)$ together with Eqs.~(\ref{CGF1}) and (\ref{CGF2}) we can identify $\rho_n(t)$ as the density matrix for the system where $n$ particles have been added to the drain during time $t$ (i.e. ${\rm Tr}(\rho_n(t))=P(n,t))$.\cite{Emary2007,Flindt2010,Marcos2010,Emary2012,Kaasbjerg2015} It also holds that $\sum_n \rho_n(t)=\rho(t)$ where $\rho(t)\equiv \rho(\chi=0;t)$ is the usual system density matrix.

To find $\rho(\chi;t)$, we use a master equation approach to deal with relaxation and dissipation, treating the coupling between the system and bath as a perturbation. Equation~(\ref{eq:rhoDynamicsWithCountingField}) yields the von Neumann
equation including the counting field
\begin{subequations}
  \begin{align}
    \label{eq:VonNeumannEqWithCountingField}
    \dot\rho_\tot(\chi;t)&=-i\calL_H(\chi)\rho_\tot(\chi;t)\\
    \calL_H(\chi)X&=\frac{1}{\hbar}(H(\chi/2)X-XH(-\chi/2)).
  \end{align}
\end{subequations}
The von Neumann equation Eq.~\eqref{eq:VonNeumannEqWithCountingField} can be
used to derive the master equation. The Hamiltonian is split into a part acting
on the system part of the Hilbert space $H_S$, a part acting on the bath part of
the Hilbert space $H_B$ and an interaction Hamiltonian $H_I$ coupling those two parts. 
As neither $H_S$ nor $H_B$ change the number of electrons in the bath we
find that
\begin{equation}
  \label{eq:DefHamiltonianWithCountingFields}
  H(\chi)=H_S+H_B+H_I(\chi).
\end{equation}
The coupling thus is the only part obtaining a counting field
dependence. Keeping track of this counting field dependence one can derive a
time local master equation by using the Markov and the secular
approximation. \cite{Blum1996,Breuer2002} This time local equation can then be
written as
\begin{equation}
\label{eq:MasterEqWithZountingFieldsGeneral}
\dot\rho(\chi;t)=-\calL(\chi)\rho(\chi;t),
\end{equation}
where $\calL(\chi)$ is a Liouvillian describing the dynamics of
$\rho(\chi;t)$. The system density matrix $\rho(t)$ is obtained via ${\cal L}(0)$ and therefore  Eq.~(\ref{eq:MasterEqWithZountingFieldsGeneral}) does not preserve the trace of $\rho(\chi;t)$ for $\chi\neq 0$.

Usually one is interested in the properties of the CGF for long measurement
times. This long time limit can be obtained from the spectral properties of
$\calL(\chi)$. \cite{Bagrets2003} The formal solution to the master equation can
be expressed by a matrix exponential
\begin{equation}
  \label{eq:FormalSolutionMasterEquation}
  \rho(\chi;t)=e^{-\calL(\chi) t}\rho(\chi;0).
\end{equation}
In this form, the solution is a sum of exponentially decaying terms,
where the timescale of the decay is given by the real parts of the eigenvalues
of $\calL(\chi)$. The long time behavior of the CGF is
determined by the eigenvalue with the smallest real part as this eigenvalue
will dominate the long time behavior. For $\chi=0$, we know that $\calL(\chi)$
reproduces the standard master equation and thus that one eigenvalue is zero
leading to the steady state whereas the other eigenvalues have a positive real
part corresponding to dephasing or relaxation. As the eigenvalues behave smoothly
when varying $\chi$ we know that the eigenvalue $\Lambda_0(\chi)$, defined as the eigenvalue that is smoothly connected to the eigenvalue $0$ for $\chi=0$,
has the smallest real part. After a sufficient long time the behavior of the CGF
will thus be dominated by $\Lambda_0(\chi)$ such that
\cite{Bagrets2003}

\begin{equation}
  \label{eq:CGFInLongTimeLimit}
  S(\chi;t)=-\ln\big(\Tr(\rho(\chi;t))\big)\approx\Lambda_0(\chi)t.
\end{equation}
To determine the cumulants the eigenvalue $\Lambda_0(\chi)$ can be expanded in
$\chi$ and the expansion coefficients can be determined using
Rayleigh-Schrödinger perturbation theory.\cite{Flindt2010} As $\calL(\chi)$ is
non-Hermitian we have to distinguish between left and right eigenvectors. Because
$\calL(0)$ determines the master equation the left and the right eigenvector for the
eigenvalue $0$ can be obtained from physical arguments; The right eigenvector is
the steady state as it has no dynamics and the left eigenvector is the linear
form defined by the trace as the trace of the density matrix is conserved in a
master equation.\cite{Breuer2002} These left and right eigenvectors are denoted
by $\langle\tilde\phi_0|$ and $|\phi_0\rangle$, where the tilde indicates the left
eigenvector. It is convenient to define
$\langle\langle\bullet\rangle\rangle\equiv\langle\tilde\phi_0|\bullet|\phi_0\rangle=\Tr(\bullet\bar\rho)$,
where $\bar\rho$ is the steady state. Using this notation and by expanding
$\calL(\chi)$ and $\Lambda_0$ as
\begin{subequations}
  \begin{align}
    \label{eq:ExpansionLandLambda0}
    \calL(\chi)&=\calL(0)+\calL'(0)\chi+\calL''(0)\frac{\chi^2}{2}+\calO(\chi^3)\\
    \Lambda_0(\chi)&=\Lambda_0(0)+\Lambda_0'(0)\chi+\Lambda_0''(0)\frac{\chi^2}{2}+\calO(\chi^3),
  \end{align}
\end{subequations}
the perturbative expansion is given by\cite{Flindt2010}
\begin{subequations}
  \begin{align}
    \label{eq:LabmdaZeroExpansionCoeffsInPert}
    \Lambda_0'(0)&=\langle\langle\calL'\rangle\rangle\\
    \Lambda_0''(0)&=\langle\langle\calL''\rangle\rangle
      -\langle\langle\calL'\calQ\frac{2}{\calL(0)}\calQ\calL'\rangle\rangle,
  \end{align}
\end{subequations}
where $\calP=|\tilde\phi_0\rangle\langle\phi_0|$ and $\calQ=\idop-\calP$
are the projectors onto the eigenspace of $\Lambda_0(0)$ and its complement and
the prime denotes the derivative with respect to the counting field at $\chi=0$. Using these
expansion coefficients the mean current and the current noise are given by
\begin{subequations}
  \begin{align}
    \bar I&=i\frac{e}{{\cal T}}\frac{\partial}{\partial\chi}S(\chi,{\cal T})|_{\chi=0}
    =i e \Lambda_0'(0)\\
    S&=\frac{e^2}{{\cal T}}\frac{\partial^2}{\partial\chi^2}S(\chi,{\cal T})|_{\chi=0}
    =e^2\Lambda_0''(0),
  \end{align}
\end{subequations}
where ${\cal T}$ is a large measurement time during which the electrons are counted. Thus, using the spectral properties of $\calL(\chi)$ we were able to calculate the
transport properties.

\subsection{Jump operators and coincidence correlation functions}

\label{sec:jump-operators}

The counting statistics approach can be extended to describe also the
joint probability of the outcome of multiple
measurements. \cite{Emary2007} This enables discussion of bunching and
antibunching of electron transfer events within the same formalism.
\cite{Emary2012}

Let us define the generating function for
the joint probability of observing $n_j$ particles having been
transmitted during time intervals $[t_j,t_{j-1}]$ for $j=1,\ldots,M$:
\begin{align}
 \label{eq:Smulti}
 e^{-S(\vec{\chi},\vec{t})}
 =
 \sum_{n_1,\ldots,n_M}
 e^{i\sum_{j=1}^Mn_j\chi_j}
 P(n_M,t_M;\ldots;n_1,t_1)
 \,,
\end{align}
where $\vec{\chi}=(\chi_1,\ldots,\chi_M)$, $\vec{t}=(t_1,\ldots,t_M)$
and $t_0\equiv0$.  Under the Markovianity assumptions,
the outcome $n_j$ depends on the earlier outcomes $n_i$, $i<j$, only
via the state of the system. Hence, the generating function in
Eq.~\eqref{eq:Smulti} can be evaluated in the same way as done for a
single measurement in the previous subsection, but taking the initial
state for the time evolution \eqref{eq:MasterEqWithZountingFieldsGeneral}
on each interval to be the appropriate conditional density matrix,
conditioned on the previous measurement outcomes.  Then, as discussed
in Ref.~\onlinecite{Emary2007},
Eq.~\eqref{eq:FormalSolutionMasterEquation} generalizes to
\begin{align}
 \rho(\vec{\chi},\vec{t})
 =
 e^{-\mathcal{L}(\chi_M)(t_M-t_{M-1})}\cdots{}e^{-\mathcal{L}(\chi_1)(t_1-0)}\rho_0
 \,,
\end{align}
and $\exp(-S(\vec{\chi},\vec{t}))=\Tr[\rho(\vec{\chi},\vec{t})]$.  That is,
$\rho(\vec{\chi},\vec{t})$ follows the same master
equation~\eqref{eq:MasterEqWithZountingFieldsGeneral} as for the
single-measurement case, but with $\chi$ replaced by $\chi_j$ for
$t_j>t>t_{j-1}$.

The delayed coincidence (``Glauber'') correlation functions can be
determined in the above approach from the electron counting
coincidence rates. \cite{Emary2012} Consider the joint probability
$G_2$ of observing ``clicks'' i.e. electron transfers $\mu,\nu$ (which
can be any integers) during both time intervals $[\tau+\delta,\tau]$
and $[\delta,0]$ for $\tau>\delta$, and similarly the probability
$G_1$ of observing the events separately, when the system is initially
in the steady state, i.e. $\rho_0={\bar \rho}$. They can be expressed
as
\begin{align}
 G_{2,\mu\nu}
 &=
 \sum_{n}P(\mu,\tau+\delta;n,\tau;\nu,\delta)
 \\
 &=
 \int_0^{2\pi}\frac{d\chi_1\,d\chi_2}{(2\pi)^2}
 e^{-i\mu\chi_1-i\nu\chi_2}
 \Tr[
 e^{-\delta \mathcal{L}(\chi_1)}
 \\&\notag
 \qquad\times
 e^{-(\tau-\delta)\mathcal{L}(0)}
 e^{-\delta \mathcal{L}(\chi_2)}
 {\bar \rho}
 ]
 \,,
 \\
 G_{1,\mu}
 &=
 P(\mu,\delta)
 =
 \int_0^{2\pi}\frac{d\chi}{2\pi}
 e^{-i\mu\chi}
 \Tr[
   e^{-\delta \mathcal{L}(\chi)}
   {\bar \rho}
 ],
\end{align}
where we sum over $n$ as we ignore the number of electrons (clicks)
during $[\tau,\delta]$. For $\delta\to0$, it then follows that
\begin{align}
 G_{2,\mu\nu}
 &\simeq
 \delta^2\langle\langle  \mathcal{J}_\mu e^{-\tau\mathcal{L}(0)}  \mathcal{J}_\nu \rangle\rangle
 \,,
 \\
 G_{1,\mu}
 &\simeq
 -\delta\langle\langle  \mathcal{J}_\mu \rangle\rangle
 \,,
\end{align}
where $\mu,\nu\ne0$, and
\begin{align}
 \mathcal{J}_\mu = \int_0^{2\pi}\frac{d\chi}{2\pi}e^{-i\mu\chi}\mathcal{L}(\chi)
 \label{jumpoperator1}
\end{align}
are the jump superoperators.  Indeed, the master
equation~\eqref{eq:MasterEqWithZountingFieldsGeneral} can be written in terms of the
number-resolved density matrices as
\begin{align}
 \dot{\rho}(n,t)
 &=
 \sum_{\nu=-\infty}^\infty \mathcal{J}_\nu \rho(n-\nu,t)
 \,,
\end{align}
where each $\mathcal{J}_\nu$ describes processes transferring $\nu$ electrons.

The delayed coincidence correlation functions can now be defined as the ratio of
the actual joint probability to that of uncorrelated events:
\begin{align}
 g^{(2)}_{\mu\nu}(\tau)
 =
 \frac{G_{2,\mu\nu}}{G_{1,\mu}G_{1,\nu}}
 =
 \frac{
   \langle\langle \calJ_\mu e^{-\tau\mathcal{L}(0)} \calJ_\nu \rangle\rangle
 }{
   \langle\langle \calJ_\mu \rangle\rangle
   \langle\langle \calJ_\nu \rangle\rangle
 }
 \,,
\end{align}
for $\delta\to0$. Note that here we allow for events of different
types $(\mu,\nu)$ to occur.

The behavior of this function for $\tau\rightarrow 0$ can be used to
characterize the bunching behavior of the events. For $g^{(2)}_ {\mu\nu}(0)>1$
the two events tend to follow each other more likely than they would if they were
independent and thus bunch. For $g^{(2)}_{\mu\nu}(0)<1$ the events repel each
other and thus are antibunched.

To connect the noise to the generalized $g^{(2)}$
function we define the reduced propagator
$\calR(\tau)\equiv\calQ\exp(-\calL(0)\tau)\calQ$ which allows us to write the
$g^{(2)}_{\mu\nu}$ function as
\begin{equation}
\label{eq:DefGeng2FuncRedProp}
g^{(2)}_{\mu\nu}(\tau)
=
1+\frac{
\langle\langle\calJ_\mu \calR(\tau)\calJ_\nu\rangle\rangle
}{
\langle\langle\calJ_\mu \rangle\rangle
\langle\langle\calJ_\nu \rangle\rangle
},
\end{equation}
where we used that $\calP\calQ=\calQ\calP=0$. By integrating the reduced
propagator we obtain the pseudo inverse
\begin{equation}
\label{eq:ConnectionPsudiInvReProp}
\calQ\frac{1}{\calL(0)}\calQ=\int_0^\infty\calR(\tau)\dd\tau,
\end{equation}
needed in the perturbative expansion. 

Using the inverse Fourier transform of Eq.~(\ref{jumpoperator1}), we obtain $\calL'(0)=\sum_{\nu} i\nu \mathcal{J}_\nu$ and
$\calL''(0)=-\sum_{\nu}\nu^2\mathcal{J}_\nu$ where $\nu$ runs over all integers. This together with the representation of the pseudo inverse
we insert into the perturbative expansion of the average backscattering current and the current noise and obtain
\begin{equation}
{\bar I}=\sum_{\nu} I_{\nu},
\label{noiseGlauber}
\end{equation}
and 
\begin{multline}
S=e\sum_{\nu}\nu I_{\nu}+2\sum_{\mu,\nu}I_{\mu}I_{\nu} \int_{0}^{\infty} d\tau\left(g^{(2)}_{\mu\nu}(\tau)-1\right),
\label{eq:FanoByG2Function}
\end{multline}
where we have defined $I_{\nu}=-e\nu \langle\langle\mathcal{J}_{\nu}\rangle\rangle$.

The first term in Eq.~(\ref{eq:FanoByG2Function}) directly corresponds to noise due to independent scattering events. If transport is dominated by a single process (i.e. only one $\nu$ contributes), this term describes Poissonian noise since the Fano factor $F=S/e|{\bar I}|$ becomes in this case $F=|\nu|$ which is directly proportional to the fundamental 
charge transported in an event (e.g. $|\nu|=1$ for a single electron process or $|\nu|=2$ for simultaneous transport of two electrons like for Cooper pairs). 
The second term in Eq.~(\ref{eq:FanoByG2Function}) shows the genuine correlations between subsequent scattering events via the integrated coincidence functions $g^{(2)}_{\mu\nu}(\tau)$. If coincidences are absent $g^{(2)}_{\mu\nu}(\tau)=1$, and the second term vanishes. However, in many problems (see the one we analyze in the next section) more than one process contributes. Eq.~(\ref{eq:FanoByG2Function}) then shows that a simple connection of noise or Fano factor to the bunching or anti-bunching behavior of subsequent scattering processes is not possible due to the summations over $\mu,\nu$. This is in contrast to the analysis in Ref.~\onlinecite{Emary2012}, where only one kind of an event was possible.

Equations~(\ref{noiseGlauber}) and (\ref{eq:FanoByG2Function}) are the main results of this section. They present the average backscattering current and current noise in terms of jump superoperators, evaluated in the steady state, that describe the evolution of the reduced density matrix due to charge transfer events ("jumps") between the reservoir and the system. The current noise Eq.~(\ref{eq:FanoByG2Function}) is directly related to the coincidence functions taking into account all possible
kinds of charge transfers. The only assumptions used in the derivation is a Lindblad form of the master equation (Eq.~(\ref{eq:MasterEqWithZountingFieldsGeneral})) and a CGF that can be expanded in the counting field. Under these premises, the results are general and valid for all temperatures and bias voltages applied to the
reservoir(s).

\section{Spin $1/2$ impurity in a magnetic field coupled to a helical edge}
\label{sec:Spin12Impurity}

Transport in helical Luttinger liquids is sensitive to the coupling to magnetic impurities \cite{Maciejko2009, Tanaka2011, Posske2013, Altshuler2013, Nagaev2018, Kurilovich2019, Pashinsky2020}, and in particular, certain details of the dynamics of the impurities become visible in the transport current through the edge.
To study this, we apply the formalism derived in the last section to a Zeeman split single level quantum dot (QD) in the cotunneling
regime tunnel coupled to a helical Luttinger liquid (HLL), as illustrated in
Fig.~\ref{fig:setup}.

\subsection{The model}

In the cotunneling regime, the coupling to the edge state
can be described by the Kondo Hamiltonian \cite{Hewson1993}. The Hamiltonian of
the total system is given by
\begin{equation}
  H=H_\HLL+H_Z+H_K,
\end{equation}
where $H_\HLL$ describes the edge state electrons, $H_Z$ the effect of the
magnetic field on the QD and $H_K$ the coupling of the QD to the edge states.

The electrons in the helical edge are forming a helical Luttinger liquid (HLL)
described by
\begin{align}
  H_\HLL
  &=\hbar \vF\int\dd{\xi}:\sum_{\eta=\pm}
\Psi_\eta^\dagger(\xi)(-i\eta\partial_\xi)\Psi_\eta(\xi):
  \\\notag
  &\quad+\frac{\lambda}{2}\int 
\dd{\xi}:\Big(\sum_{\eta=\pm}\Psi_\eta^\dagger(\xi)\Psi_\eta(\xi)\Big)^2:,
  \label{eq:DefHHLL} 
\end{align}
where $\vF$ is the Fermi velocity, $\Psi^{(\dagger)}_\eta(\xi)$ is the electron
field operator on the branch $\eta=\pm$, $\lambda$ is the strength of the Coulomb
repulsion and $:\bullet:$ denotes normal ordering.
By applying a bias voltage $V$ to
the edge states (see Fig.~\ref{fig:setup}), right $(+)$ and left $(-)$ movers acquire different chemical potentials $\mu_{\pm}=\pm eV/2$
(electron charge is $-e=-|e|$). The chemical potential can be gauged into the field operators
$\Psi_{\pm}\mapsto\exp(-i\mu_{\pm}t)\Psi_{\pm}$ where $t$ denotes time \cite{Peca2003}.   

In the cotunneling regime the coupling of the quantum dot spin to the edge states can be described by an isotropic Kondo
Hamiltonian,\cite{Hewson1993} which can be written as
\begin{equation}
  \label{eq:DefHkondo}
  H_K = 2 J \bm{S}_L\cdot\bm{S}_{QD},
\end{equation}
where $2J$ is the Kondo coupling strength. The spin operators are defined by
$S_{L}^k=\tfrac{\hbar}{2}\sum_{\mu,\nu}\Psi_\nu^\dagger(0)\hat{\sigma}^k_{\nu\mu}
\Psi_\mu(0)$ and
$S_{QD}^k=\tfrac{\hbar}{2}\sum_{\mu,\nu}d_\nu^\dagger\hat{\sigma}^k_{\nu\mu}d_\mu$,
where $\hat{\sigma}^k$, $k=x,y,z$, are the Pauli matrices and $d^{(\dagger)}_\nu$
are the operators for electrons with spin $\nu$ on the QD.  The ladder operators are defined
as $S_L^\pm=S_L^x\pm iS_L^y$ and $S_{QD}^\pm=S_{QD}^x\pm iS_{QD}^y$. Here we
used the spin quantization axis of the electrons in the edge states such that
the spin of the electrons created by $d_\nu^\dagger$ is defined with respect to this axis.

The magnetic field on the QD shall point in an arbitrary direction ${\hat n}=(\sin\theta_Z\cos\phi,\sin\theta_Z\sin\phi,\cos\theta_Z)^{T}$
such that the Zeeman effect is described by
\begin{equation}
  \label{eq:DefHzeeman}
  H_Z = g\mu_B \bm{B}\cdot\bm{S},
\end{equation}
where $g$ is the $g$ factor, $\mu_B$ is the Bohr magneton, and
$\Delta_Z=g\mu_B|\bm{B}|$ is the resulting Zeeman splitting\footnote{For
  definiteness, we assume $g>0$ in the paper. The case $g<0$ can be accounted
  for by reversing the sign of ${\bm B}$}. We note that we envision the situation where a Zeeman field
 is present only in the QD, whereas the spin quantization axis in the helical edge state leads is unaffected by the Zeeman-field.
 This situation is either realized by restricting the magnetic field to the QD, or by having a global magnetic field but a large Fermi energy (compared $\Delta_Z$) in the leads \cite{Probst2015}.
The coupling to the edge state might
lead to an induced magnetic field which would add another term to $H_Z$ \cite{Probst2015}. 
Here, we restrict ourselves to a parameter regime
$JV/(4\pi v_F)\ll\Delta_Z$ in which this term can be neglected \cite{Probst2015}. To diagonalize $H_Z$, we choose a new spin quantization
axis $" ' "$ along to the magnetic field. This is effectively a rotation of the spin
quantization axis ${\hat z}$ where $\theta_Z$ is the angle of the tilt of the $z$ axis and
$\phi$ determines the direction of this tilt. Including the final rotation along
the resulting $z$ axis $\gamma$ the spin operators $S_{QD}^i$ can be expressed
by spin operators defined with respect to the tilted spin quantization axis
$S'^i_{QD}$ by $S^{i}_{QD}=\sum_{ij=\pm,z}\mathscr{D}(U)_{ij}{{S'}^j_{QD}}$,
where, written in the basis $(S^z,S^+,S^-)$,
\begin{equation}
  \label{eq:OrthogonalRotOfSpin}
  \mathscr{D}(U)=
    \begin{pmatrix}
\cos\theta_Z&-\tfrac{z_\gamma}{2}\sin\theta_Z&-\tfrac{z_\gamma^*}{2}\sin\theta_Z\\
      z_\phi\sin\theta_Z&z_\phi 
z_\gamma\cos^2\frac{\theta_Z}{2}&-z_\gamma^*z_\phi\sin^2\frac{\theta_Z}{2}\\
      z_\phi^*\sin\theta_Z&-z_\gamma 
z_\phi^*\sin^2\frac{\theta_Z}{2}&z_\gamma^*z_\phi^*\cos^2\frac{\theta_Z}{2} 
      \end{pmatrix},
\end{equation}
using $z_\phi\equiv e^{i\phi}$ and $z_\gamma\equiv e^{i\gamma}$. The
coefficients of this transformation are defined as
$c_{kl}\equiv(\mathscr{D}(U))_{kl}$.

\subsection{Master equation}

We now derive the master equation using the formalism from Section I. The number of left-movers ${\hat N}_-$ is considered the particle number to be counted, and associated with the counting field $\chi$. Since the coupling $H_K$ conserves the total particle number, the change in $N_-$ corresponds directly to the backscattering current.

We first calculate $H_I(\chi)=\exp(i\chi{\hat N}_-)H_K\exp(-i\chi{\hat N}_-)$. For convenience, we define new operators\footnote{Including the induced field into
  Eq.~(\ref{eq:DefHzeeman}) would alter the definition of $A_z$ as explained in
  Ref.~\citenum{Probst2015}.} $A_\pm=J S_L^\mp$ and $A_z=2J S_L^z$ such that
\begin{equation}
  \label{eq:HIntWithCountingField}
  H_I(\chi)=\sum_{k=\pm,z}e^{i\sigma_k\chi}A_k S^k_{QD},
\end{equation}
where $\sigma_+=-\sigma_-=1$ and $\sigma_z=0$. With this and
Eq.~(\ref{eq:VonNeumannEqWithCountingField}) we follow the derivation of the
master equation using standard steps \cite{Breuer2002,Probst2015}.
According to Eq.~(\ref{eq:DefHamiltonianWithCountingFields}), we identify the bath Hamiltonian $H_B$ with $H_{\rm HLL}$ and the system Hamiltonian $H_S$ with $H_Z$
and obtain 
\begin{equation}
{\dot \rho}^{I}(\chi;t)=-\int_{t_0}^{t} d\tau{\rm Tr}_B\left({\cal L}_{I}^{I}(\chi;t){\cal L}_{I}^{I}(\chi;\tau)\rho_{\rm tot}^{I}(\chi;\tau)\right).
\label{MEQ1}
\end{equation}
Here, the superscript index $I$ denotes the interaction picture with respect to $H_0=H_{\rm HLL}+H_Z$ and ${\cal L}_{I}^{I}(\chi;t)$ is the Liouvillian with respect to $H_{I}^{I}(\chi)$. The result can be transformed back to the Schrödinger picture by adding a
commutator with the system Hamiltonian. By associating the phase factors in
Eq.~\eqref{eq:HIntWithCountingField} with the $S_{QD}^k$ operators the counting field
phase factors can easily be added to the master equation, where the sign in the
exponent depends on whether the operator is put to the left or to the the right of the reduced
density matrix $\rho^{I}(\chi;t)$. By performing the Markov- and secular approximation \cite{Breuer2002,Probst2015} and by including the phase factors in the bath correlation functions and employing the eigenbasis of $H_S$ we find for the master
equation
\begin{multline}
  \label{eq:QDHLLFCSMasterEquationSecularEithCountingFieldInteractionPicture}
  \dot\rho(\chi;t)=-\frac{i}{\hbar}[H_Z,\rho(\chi;t)]+
  \frac{1}{\hbar^2}\sum_{k=\pm,z}\\
  \Big(
  \calF_{\bar kk}(\chi;-\Delta_Z\sigma_k)S'^k_{QD}\rho(\chi;t)(S'^k_{QD})^\dagger\\
  -\frac{1}{2}
  \calF_{\bar k k}(0;-\Delta_Z\sigma_k)
  \lbrace (S'^k_{QD})^\dagger S'^k_{QD},\rho(\chi;t)\rbrace
  \Big),
\end{multline}
where $\calF_{k \bar k}(\chi;\omega)=\sum_{\alpha=\pm,z}
e^{-i\sigma_\alpha\chi}c_{\alpha k}c_{\bar\alpha \bar k}
F_{\alpha\bar\alpha}(\omega)$ is given by the Fourier transform of the lead
correlation functions
$F_{\alpha\bar\alpha}(\omega)=\int_{-\infty}^\infty\dd{\tau}e^{i\omega\tau}
\langle A_\alpha(\tau)A_{\bar\alpha}(0)\rangle$, $\bar k=-k$ for $k=\pm$ and $\bar z=z$. 
Above, we omitted the small Lamb shift term, which renormalizes $H_Z$.

The lead correlation functions can be calculated
using a standard bosonization approach\cite{Giamarchi2007, Probst2015}
\begin{subequations}
  \label{eq:FourTransLeadCorrFunc}
  \begin{align}
    F_{zz}(\omega)&= \frac{1}{\hbar\beta}\bigg(\frac{\hbar J}{v}\bigg)^2
                  \frac{\omega\beta}{2 \pi K}
                 \frac{e^{\beta\omega/2}}{\sinh(\omega\beta/2)}
                 \\
    F_{\alpha\bar\alpha}(\omega)&
               =\frac{1}{\hbar\beta}\bigg(\frac{\hbar J}{v}\bigg)^2
               (2 a)^{2K-2}\nonumber\\
               &\times
               \frac{\pi}
               {\Gamma(2K)|\Gamma(1-K+i(\omega-\sigma_\alpha e V)\beta/2\pi)|^2}
               \nonumber\\
               &\times
               \frac{e^{(\omega-\sigma_\alpha e V)\beta/2}}
               {\cosh\big((\omega-\sigma_\alpha e V)\beta\big)
               -\cos\big( 2\pi K\big)},
  \end{align}
\end{subequations}
where $a\equiv\pi\alpha/\hbar\beta v$, $\alpha$ is the short distance cutoff, $\beta=1/k_BT$
and $K\equiv1/\sqrt{1+\lambda/\pi\hbar\vF}$ is the interaction parameter. We also defined the velocity of charge excitations $v=v_F/K$.
In secular approximation diagonal and off-diagonal entries decouple and can be
treated separately. The off-diagonal entries of the density matrix decay due to
dephasing such that it is sufficient to discuss the dynamics of the diagonal
elements. For long times $t$, the reduced density matrix can then be represented by a vector
$\rho(\chi;t)=(\rho_{\up}(\chi;t)\ \rho_{\down}(\chi;t))^T$. In this representation, the master equation Eq.~(\ref{eq:MasterEqWithZountingFieldsGeneral})
can be represented by a Liouvillian ${\cal L}(\chi)={\cal L}_0+e^{i\chi}{\cal J}_+ +e^{-i\chi}{\cal J}_{-}$ with
\begin{subequations}
\begin{align}
  \label{eq:DefDecompJumOpByRates}
  \mathcal{J}_\pm&=
  \begin{pmatrix}
    -\Gamma_{00}^\pm&-\Gamma_{\up\down}^\pm\\
    -\Gamma_{\down\up}^\pm&-\Gamma_{00}^\pm
  \end{pmatrix}\\
  \mathcal{L}_0&=
  \begin{pmatrix}
    \Gamma_{\down\up}+\Gamma_{00}^0&-\Gamma_{\up\down}^0\\
    -\Gamma_{\down\up}^0&\Gamma_{\up\down}+\Gamma_{00}^0
  \end{pmatrix},
\end{align}
\end{subequations}
where the rates are given by
\begin{subequations}
  \label{eq:DefOfRates}
  \begin{align}
    \Gamma_{00}^0 &= \frac{1}{4}\sin^2\theta \ \mathrlap{\Big(F_{+-}(0)+F_{-+}(0)\Big)
    \,,}
    \\
    \Gamma_{00}^+ &= \frac{1}{4}\sin^2\theta \ F_{-+}(0)
    \,,
    &
    \Gamma_{00}^-&=\frac{1}{4}\sin^2\theta \ F_{+-}(0)
    \,,
    \\
    \Gamma_{\up\down}^0 &= \frac{1}{4}\sin^2\theta \ F_{zz}(-\Delta_Z)
    \,,
    &
    \Gamma_{\down\up}^0 &= \frac{1}{4}\sin^2\theta \ F_{zz}(\Delta_Z)
    \,,
    \\
    \Gamma_{\up\down}^+ &=
    \cos^4\frac{\theta}{2} \ F_{-+}(-\Delta_Z)
    \,,
    &
    \Gamma_{\down\up}^+ & = \sin^4\frac{\theta}{2} \ F_{-+}(\Delta_Z)
    \,,
    \\
    \Gamma_{\up\down}^- &= \sin^4\frac{\theta}{2} \ F_{+-}(-\Delta_Z)
    \,,
    &
    \Gamma_{\down\up}^- &= \cos^4\frac{\theta}{2} \ F_{+-}(\Delta_Z)
    \,,
    \\
    \Gamma_{\sigma\bar\sigma} &= \Gamma^0_{\sigma\bar\sigma} + \Gamma^+_{\sigma\bar\sigma} + \Gamma^-_{\sigma\bar\sigma}
    \,.
  \end{align}
\end{subequations}
The superscript denotes whether the rate corresponds to a
process that changed the number of left movers whereas the subscript denotes the
process associated with the QD. A zero indicates that the process does not change
the state. A superscript $0$ thus corresponds to processes on the QD without a
change in the edge channel whereas a $00$ subscript indicates that a cotunneling
process without a change of the QD state has occurred. All rates with a nonzero
subscript are thus inelastic cotunneling rates ($\Delta_Z >0$), all other processes correspond to elastic cotunneling rates.
Note that the dependence on $\Gamma_{00}^{\eta}$, $\eta=0,+,-$ disappears in ${\cal L}(\chi)$ in the case $\chi=0$.
This is easily understood by recalling that the case $\chi=0$ describes the master equation for the system density matrix $\rho(t)$ which only contains rates that
change the state of the system (i.e. the QD). The total rate for a specific change of the state of the QD is
denoted by $\Gamma_{\sigma\bar\sigma}$.
The steady state ${\bar \rho}$ of the QD satisfies ${\cal L}(0){\bar \rho}=0$ which leads to
\begin{equation}
  \label{eq:DefSteadyStateByRates}
  {\bar \rho}=\frac{1}{\Gamma}
  \begin{pmatrix}
    \Gamma_{\up\down}\\\Gamma_{\down\up}
  \end{pmatrix},
\end{equation}
where $\Gamma\equiv\Gamma_{\up\down}+\Gamma_{\down\up}$.
The detailed behavior of the different rates is discussed in Sec.~\ref{Sec:TunnelingRates}.
In the above model, the matrices $\mathcal{J}_\pm$ have identical
diagonals, i.e., the elastic cotunneling rates do not depend on the
internal state of the QD.  If this was not the case and the elastic
rates differed by $\delta\Gamma_{00}$, slow switching of the QD state
could generate telegraphic noise in the current,
\cite{Sukhorukov2001,Kaasbjerg2015} increasing the Fano factor by
$\delta{}F\sim{}\delta\Gamma_{00}/\Gamma$ if the inelastic rates are
small compared to elastic rates.  This mechanism is here generally
absent due to the structure of the Kondo model master equation, as the
two elastic transport processes are equal as both are mediated only by
$S_{QD}^{\prime z}$. As in our case the elastic rates are not
significantly larger than the inelastic ones, the effect of including
small corrections in the model would likely be negligible.


\begin{figure}
  \centering
  \includegraphics[width=8cm]{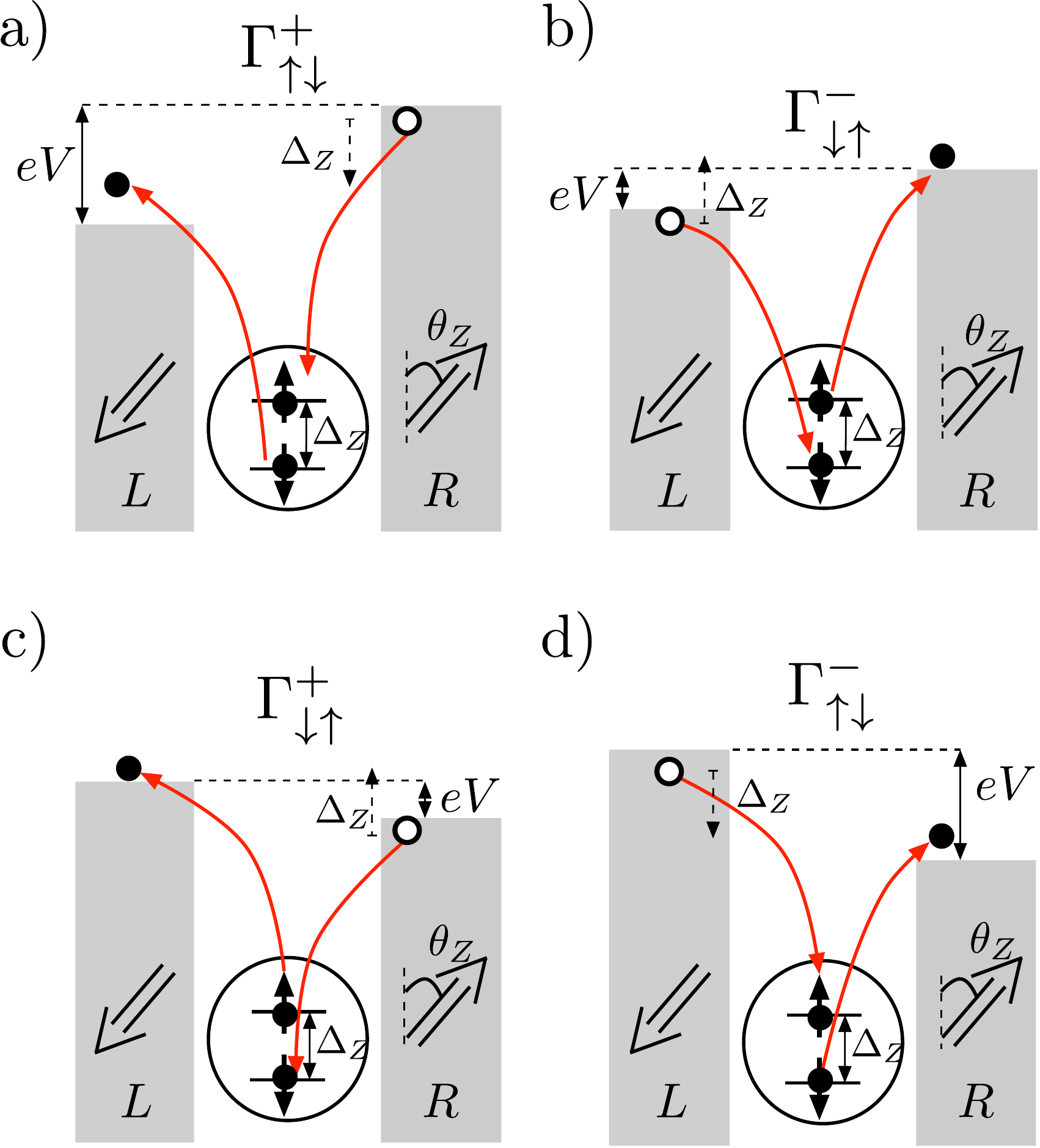}
  \caption{
    Schematic of the inelastic processes that contribute
    to the backscattering current. The processes move electrons
    either from right-movers to left-movers $(+)$ or vice versa $(-)$,
    while simultaneously flipping the dot state.
    At $\theta_Z=0$, the ``spin-conserving'' rates $\Gamma_{\uparrow\downarrow}^+$ and $\Gamma_{\downarrow\uparrow}^-$ (a) and b))
    are nonzero, whereas the ``non-spin-conserving'' rates $\Gamma_{\downarrow\uparrow}^+$ and $\Gamma_{\uparrow\downarrow}^-$ (c) and (d)) vanish.
    The $\uparrow\downarrow$ rates excite the dot
    whereas $\downarrow\uparrow$ relax it.
  }
  \label{fig:Illustrated_Rates}
\end{figure}

\begin{figure}
  \centering
  \includegraphics{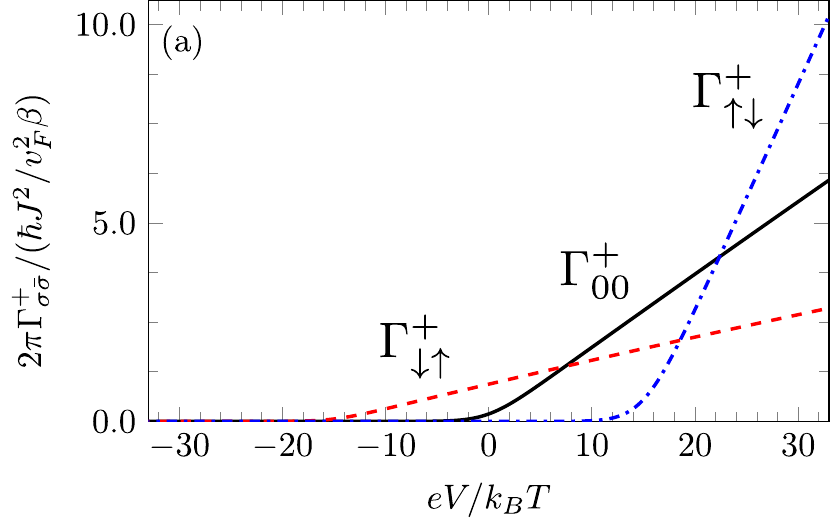}
  \includegraphics{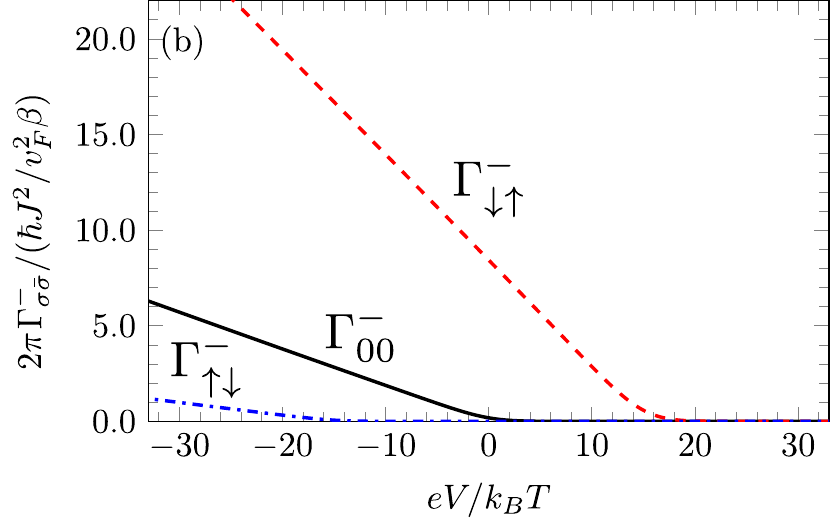}
  \caption{Rates $\Gamma^\sigma_{00}$ (black, solid), $\Gamma^\sigma_{\up\down}$
    (blue, dash-dotted) and $\Gamma^\sigma_{\down\up}$ (red, dashed) for
    $\sigma=+$ (a) and $\sigma=-$ (b)
    with parameters as in Fig.~\ref{fig:Fanodecomp}.
   }
  \label{fig:DecompOfFlipRates}
\end{figure}

\subsection{Behavior of the tunneling rates}
\label{Sec:TunnelingRates}

The behavior of the transition rates plays an important role in determining the QD
state, and the transport properties of the system, so we now discuss it first.  In the
representation of the rates Eq.~\eqref{eq:DefOfRates}, we can
associate each rate to a change of number of left-movers in the leads ($\pm,0$) and a
process on the QD (elastic $00$, relaxation $\downarrow\uparrow$, excitation $\uparrow\downarrow$).
Due to the tilt of
the spin quantization axis, electrons with a specific spin on the QD
have a finite overlap with both spin directions in the leads.  If the
spin quantization axes of the helical edge states and the QD are not
oriented perpendicular to each other, one spin direction has a larger
overlap than the other. For brevity, we call processes that couple the
spin directions with the larger overlap ``spin conserving'' and the other
processes ``spin conservation violating''.
The rate of spin conservation violating processes decreases to zero
as the orientation approaches the parallel orientation ($\theta_Z\to0$).
For perpendicular orientations both types of processes are equally strong.

Because the QD energy level is Zeeman split, spin flip processes on the
QD correspond to relaxation or excitation processes. To relax (excite)
the QD, energy needs to be deposited in (or absorbed from) the
leads. When flipping the spin of a single electron in the edge state by
reverting its propagation direction, its energy thus also needs to be
changed. Because the bias voltage applied to the edge state induces a
spin bias, it can selectively suppress certain processes. The processes
corresponding to the rates that change the number of left- and right
movers in the leads are depicted in Fig.~\ref{fig:Illustrated_Rates}
(for the inelastic processes). Their rates as a function of the bias voltage
are shown in Fig.~\ref{fig:DecompOfFlipRates} for $\theta_Z=\pi/3$.

The spin conserving rates ($\Gamma^+_{\up\down}$ and
$\Gamma_{\down\up}^-$) increase and decrease the fastest as a function of the bias voltage. The spin conservation violating rates ($\Gamma^-_{\up\down}$
and $\Gamma_{\down\up}^+$) have shallower slopes due to the smaller spin overlap. As the elastic rates $\Gamma_{00}^\pm$ do not change the state of
the QD, no energy exchange with the leads is needed, and the onset of these backscattering processes is always at $eV\approx0$. For the other rates, the onset is at $eV\approx\pm\Delta_Z$. As illustrated in Fig.~\ref{fig:Illustrated_Rates}, for $eV>\Delta_Z$ on the one hand it is
energetically possible to absorb $\Delta_Z$ from the leads by scattering a right mover
to a left mover and thereby exciting the QD spin (with rate $\Gamma^+_{\up\down}$). For $eV<\Delta_Z$, on the other hand, an energy $\Delta_Z$ can be
given to the leads by scattering a left mover to a right mover and thereby relaxing the QD spin (with rate $\Gamma_{\down\up}^-$). Both of these processes
are spin conserving. A similar argument shows that the spin conservation
violating processes have their onset at $eV\approx-\Delta_Z$. One consequence of the above is
that exciting the QD is suppressed for $|eV|<\Delta_Z$ but relaxation is possible for
all bias voltages. Moreover, importantly, for $\Gamma_{\sigma\bar\sigma}^-$ we find that relaxation is
always the dominant process whereas for $\Gamma_{\sigma\bar\sigma}^+$ the
dominant process changes as a function of the bias voltage.

\begin{figure}
  \centering
  \includegraphics{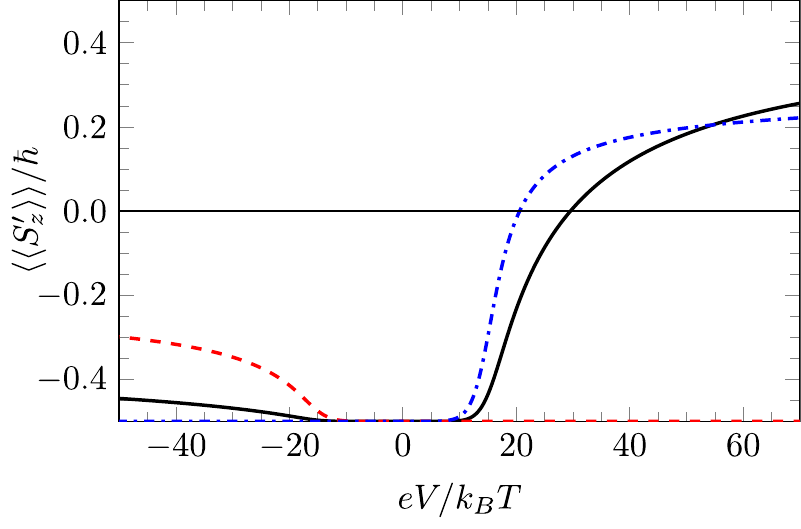}
  \caption{Polarization of the impurity spin in the steady state (black, solid)
    and polarization after a
    scattering event increasing the number of left movers (blue, dash-dotted) or
    decreasing the number of left movers (red, dashed), for $\hbar J/v_F=0.1$, $\alpha/\beta\hbar v_F=10^{-3}$, $K=1$, 
    $\Delta_Z=15\kbT$ and $\theta_Z=\pi/3$.
    }
  \label{fig:PolAfterFlip}
\end{figure}

These properties of the rates are also reflected in the steady-state polarization $\langle\langle S_z^{'}\rangle\rangle/\hbar$ of the
QD which is displayed in Fig.~\ref{fig:PolAfterFlip} (black line). We find
that the QD spin relaxes into its ground state for $|eV|<\Delta_Z$ because excitation
processes are suppressed. For $|eV|>\Delta_Z$, excitation processes are allowed
such that the QD spin becomes excited. As for $eV<-\Delta_Z$, the excitation processes
are spin conservation violating, whereas they are spin conserving for $eV>\Delta_Z$. Hence, the spin
polarization is only slightly increased for $eV<-\Delta_Z$ but even
changes sign for $eV>\Delta_Z$.

We can also consider the state of the QD immediately after a scattering event.
As discussed in Sec.~\ref{sec:Model}, the electron number increase (decrease) is
associated with the $\calJ_+$ ($\calJ_-$) jump operators.
They also give the post-measurement conditional density matrices \cite{Breuer2002},
$\bar\rho^c_\pm$, via
\begin{equation}
  \label{eq:RhoAfterScatterEvent}
  \bar\rho^{c,\pm}
  =
  \frac{\calJ_\pm\bar\rho}{\langle\langle\calJ_\pm\rangle\rangle},
\end{equation}
where it is assumed that the system was in the steady state initially. The
polarization after a flip event in the edge channel is then obtained by
evaluating the expectation value with these density matrices.

The resulting polarization after a scattering event is shown in Fig.~\ref{fig:PolAfterFlip}. We
see that reducing ($\mathcal{J}_-$, red dashed) the number of left movers always increases the polarization of the
impurity on average for $eV<0$.
This process ($\Gamma_{\uparrow\downarrow}^{-}$) is spin conservation violating, and is suppressed by the spin overlap factor.
The stronger spin conserving process ($\Gamma_{\downarrow\uparrow}^{-}$) is blocked in this case, because the QD is almost fully
in the down state before the event. For increasing the number of
left movers ($\mathcal{J}_+$), the picture is different. For $eV\approx\Delta_Z$, the spin
conserving process ($\Gamma_{\uparrow\downarrow}^{+}$) increases the spin polarization of the QD. At large enough bias, the
spin polarization of the steady state before the event however becomes positive, which starts blocking this process. For high bias voltages, the spin conservation violating process ($\Gamma_{\downarrow\uparrow}^{+}$)
thus dominates again the dynamics such that a scattering event reduces the polarization of the QD on average, compared to the initial steady state.

\subsection{Fano factor and coincidence functions}

\label{sec:fanofactorsglaube}

\begin{figure}
  \centering
  \includegraphics{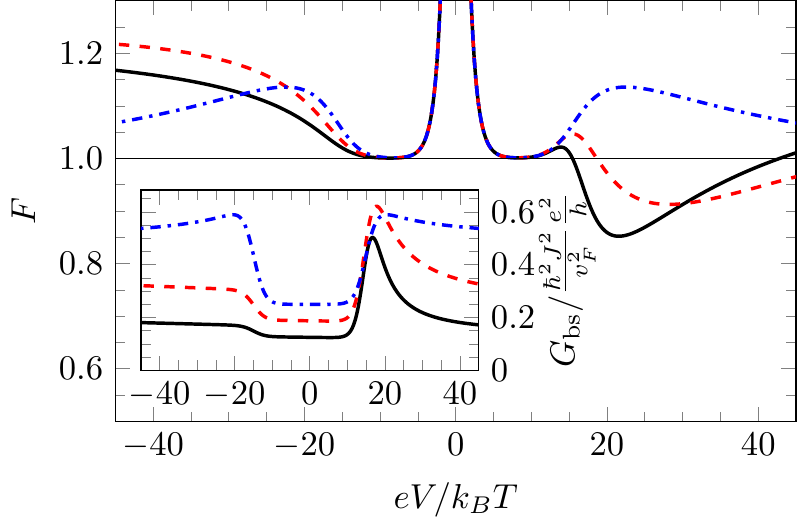}
  \caption{Fano factor for $\hbar J/v_F=0.1$, $\alpha/\beta\hbar v_F=10^{-3}$, $K=1$, $\Delta_Z=15 k_B T$,
    $\theta_Z=\pi/2$ (blue, dash-dotted), $\theta_Z=\pi/3$ (red, dashed) and
    $\theta_Z=\pi/4$ (black, solid). The inset shows the corresponding
    backscattering conductance.
    }
  \label{fig:FanoSeveralAngles}
\end{figure}

Let us now consider the backscattering current and its noise.
Using Eq.~\eqref{eq:DefDecompJumOpByRates} to evaluate
Eq.~\eqref{eq:LabmdaZeroExpansionCoeffsInPert} we are able to calculate the
backscattering current and the current noise.

In Fig.~\ref{fig:FanoSeveralAngles} the
backscattering conductance and the Fano factor are shown for several angles $\theta_Z$.
The behavior of the conductance was described in Ref.~\citenum{Probst2015}, and shows an
asymmetric onset of transport for $|eV|\approx\Delta_Z$ depending on the
relative orientation of the magnetic field and the spin quantization axis of the
electrons in the helical edge state.
For the Fano factor, we find a thermal noise divergence for
$|eV|\lesssim\kbT$, and $F\approx1$ for $\kbT<|eV|<\Delta_Z$.  In
the latter bias voltage range, elastic cotunneling processes dominate,
because excitation processes are suppressed energetically and the QD
is locked into its ground state ($|\downarrow\rangle$). In this case
the dominant processes are Poissonian elastic scattering events.
In the third regime $|eV|>\Delta_Z$, we find a
symmetric onset of super-Poissonian noise for $\theta_Z=\pi/2$. This symmetry
reflects the fact that no spin quantization axis is preferred. At $0<\theta_Z<\pi/2$,
however, the behavior is different depending on the sign of the bias voltage (c.f. e.g. $\theta_Z=\pi/3$).

\begin{figure}
\centering
\includegraphics{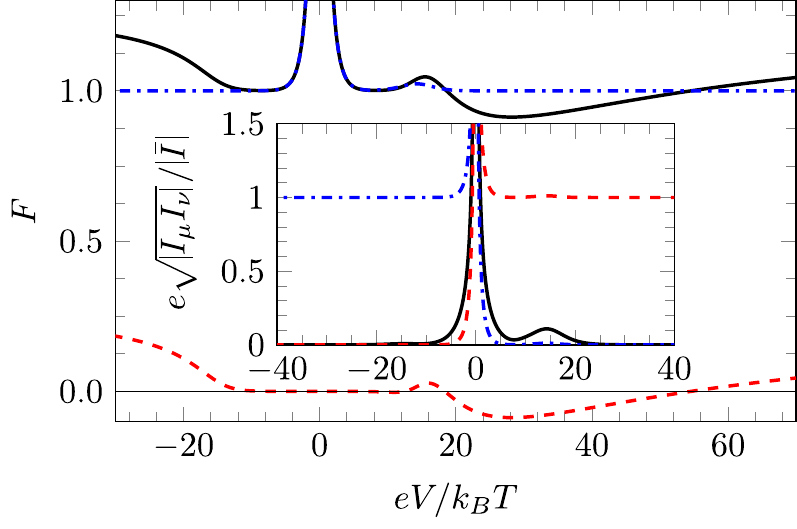}
\caption{Full Fano factor (black, solid),
  the first term (blue, dash-dotted) and the second term from Eq.~\eqref{eq:FanoByG2Function2} (red, dashed) for $\hbar
  J/v_F=0.1$, $K=1$, $\alpha/\beta\hbar v_F=10^{-3}$, $\Delta_Z=15k_BT$ and $\theta_Z=\pi/3$.
  The inset shows the weight factors
  $|I_+/\bar I|$ (red, dashed), $|I_-/\bar I|$ (blue, dash-dotted) and
  $\sqrt{|I_+I_-/\bar I|}$ (black, solid).
}
\label{fig:Fanodecomp}
\end{figure}

Before considering this behavior in more detail, we
can first have a closer look at the different terms in Eq.~\eqref{eq:FanoByG2Function}, which,
here limited to $\nu=\pm1$, gives the Fano factor as
\begin{multline}
\label{eq:FanoByG2Function2}
 F=\frac{I_+ - I_-}{|I_++I_-|}
  +\frac{2|\bar I|}{e}\int_0^\infty\dd\tau\\
 \times \bigg(
  \frac{e^2I_+^2}{\bar I^2}(g_{++}^{(2)}(\tau)-1)
  +\frac{e^2I_-^2}{\bar I^2}(g_{--}^{(2)}(\tau)-1)\\
  +\frac{e^2I_+I_-}{\bar I^2}(g_{+-}^{(2)}(\tau)+g_{-+}^{(2)}(\tau)-2)
  \bigg),
\end{multline}
where we used that $\bar I = e(I_++I_-)$ is the average backscattering current (cf.~Eq.~(\ref{noiseGlauber})).
In Fig.~\ref{fig:Fanodecomp}, we show the Fano factor as well as the first term and
the second term in Eq.~\eqref{eq:FanoByG2Function2}.
We can see that the first term, which can be attributed to independent events,
describes the thermal noise around $eV=0$ as well as the $F=1$ Poissonian noise component.
The remaining terms, which correspond to correlated events, are mainly responsible for the super- as well as the sub-Poissonian noise.
The integrand of the second term consists of three parts, each consisting of
a weight factor that determines the contribution to the current, and a
correlation factor. The weight factors are shown in the inset of
Fig.~\ref{fig:Fanodecomp}. We find that except for $eV\approx 0$, they
are dominated by two successive scattering events of the same type ($--$ for $V<0$ and $++$ for $V>0$).
Only around $eV\approx 0$ where electrons can scatter to either direction ($+-$ or $-+$), the details
are more complicated.
To understand the physics of the non-Poissonian contribution from the correlated events,
it is then sufficient to understand the $--$ and $++$ sequences of processes.

We can first consider the region $eV<-\Delta_Z$, where the noise becomes always super-Poissonian.
In general, for $eV<\Delta_Z$ the rates of relaxation processes are larger than those of
excitation processes, so that the QD is mostly in the spin down ground state, cf.~Fig.~\ref{fig:PolAfterFlip}.
For $eV<-\Delta_Z$, a (spin conservation violating) excitation process ($\Gamma_{\uparrow\downarrow}^{-}$) becomes possible, with a rate small compared to the relaxation processes (in particular $\Gamma_{\downarrow\uparrow}^{-}$, cf.~Fig.~\ref{fig:DecompOfFlipRates}),
average polarization remaining close to the spin-down state (cf.~Fig.~\ref{fig:PolAfterFlip}).
Starting from the more probable spin-down ground state, the correlated sequence of $--$ processes (i.e. both decreasing the number of left-movers in the leads) that dominates the Fano factor then consists of the slow excitation process, quickly followed by the fast relaxation process. The differing speeds of the processes now causes bunching of the two events, and results in the super-Poissonian noise.

The situation for $eV>\Delta_Z$, where also a sub-Poissonian region appears,
is more elaborate. For
$eV\approx\Delta_Z$, the spin polarization begins to increase as the excitation (with rate $\Gamma_{\uparrow\downarrow}^{+}$)
becomes faster as compared to the relaxation. This excitation
process is spin conserving, and eventually makes the
polarization after a scattering event positive (dash-dotted line in Fig.~\ref{fig:PolAfterFlip}). In this situation, a scattering event on average suppresses subsequent excitation processes. On the other hand, the relaxation (with rate $\Gamma_{\downarrow\uparrow}^{+}$) is spin conservation violating and thus has a smaller rate. In the $++$ process, a fast excitation is then preferably followed by a slow relaxation process, and we find antibunching and hence sub-Poissonian
noise. However, at yet larger voltages $eV>\Delta_Z$ the polarization of the steady state becomes positive, which starts to suppress the
excitation process. The first scattering process out of the steady state will then be preferably a spin conservation violating (i.e. lower rate) relaxation process ($\Gamma_{\downarrow\uparrow}^{+}$) which then can then be followed by the faster spin conserving excitation process ($\Gamma_{\uparrow\downarrow}^{+}$). In this case, the dominant events are bunched instead of antibunched, and the noise thus becomes super-Poissonian again.

The above discussion is indirectly affected also by processes that do not change the number of left-movers or right-movers.
Of these, the main relevant relaxation rate is $\Gamma_{\downarrow\uparrow}^{0}$, whose value is independent of the bias voltage (see Eq.~(39c)). For the parameters used in Fig.~(\ref{fig:DecompOfFlipRates}), the value is $\sim{}5.63$ in the units of the figure. Although such processes do not directly contribute to correlated transport events, they enter the steady state density matrix ${\bar \rho}$ and the polarization. As a consequence, the noise or Fano factor depends also on $\Gamma_{\downarrow\uparrow}^{0}$ and $\Gamma_{00}^{\pm}$. The rate $\Gamma_{\uparrow\downarrow}^{0}$ on the other hand is exponentially suppressed in the regime $\Delta_Z > k_B T$.

\begin{figure}
  \centering
  \includegraphics{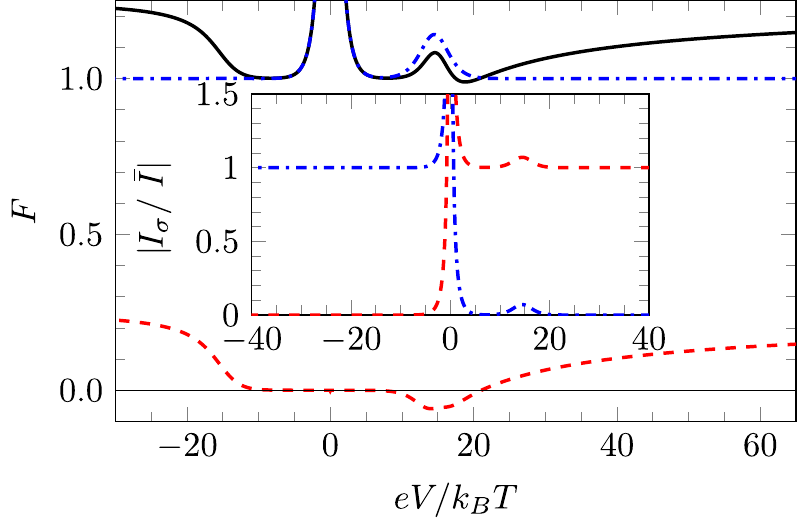}
  \caption{Fano factor (black, solid) as well as Poissonian contribution (blue,
    dash-dotted) and the contributions of correlated events for $\hbar
    J/v_F=0.1$, $\Delta_Z=15\kbT$, $\theta_Z=\pi/3$, $\alpha/\hbar\beta
    v_F=10^{-3}$ and $K=0.8$.
    The Poissonian part and contributions by
    correlations are given by the first and the second term in
    Eq.~\eqref{eq:FanoByG2Function}. In the inset the weight
    factors $|I_+/\bar I| $ (red, dashed) and $|I_-/\bar I|$ (blue, dash
    dotted) are shown.
     }
  \label{fig:FanoDecompositionWithInteraction}
\end{figure}

The above qualitative discussion can be accompanied by a more quantitative
analysis. A direct calculation shows that the eigenvalues of $\calL(0)$
are $0$ and $\Gamma$. Using these eigenvalues and that the QD has only
two states we find that
\begin{equation}
  \label{eq:PseudoInverseExplicite}
  \calQ\frac{1}{\calL(0)}\calQ=
  \frac{\calQ}{\Gamma}.
\end{equation}
Using Eq.~\eqref{eq:ConnectionPsudiInvReProp} and
Eq.~\eqref{eq:DefGeng2FuncRedProp} we find
\begin{align}
  \label{eq:IntG2MinusOneByJumpOp}
  \int_0^\infty\big(
  g^{(2)}_{\mu\nu}(\tau)-1
  \big)\dd\tau
  &=
  \frac{1}{\Gamma}
  \frac{
    \langle\langle 
    \calJ_\mu\calJ_{\nu}
    \rangle\rangle
    -
    \langle\langle 
    \calJ_\mu
    \rangle\rangle\langle\langle
    \calJ_{\nu}
    \rangle\rangle
  }{
    \langle\langle 
    \calJ_\mu
    \rangle\rangle\langle\langle
    \calJ_{\nu}
    \rangle\rangle
  }
  \\
  &=
  \frac{g^{(2)}_{\mu\nu}(0) - 1}{\Gamma}
  .
\end{align}
This illustrates that because in our case the time evolution is a simple
exponential decay, the probability of two successive scattering events determines
the bunching behavior, which determines the sign of the deviation from the Poissonian Fano factor. Inserting
Eq.~\eqref{eq:DefOfRates} and Eq.~\eqref{eq:DefDecompJumOpByRates} into
Eq.~\eqref{eq:IntG2MinusOneByJumpOp} we find
\begin{align}
  \label{eq:IntG2MinusOneByRates}
  g^{(2)}_{\mu\mu}(0) - 1
  &=
  \frac{
    \Gamma_{\up\down}^\mu\Gamma_{\down\up}^\mu
    -
    (\Gamma_{\up\down}^\mu\bar{\rho}_{\down}+\Gamma_{\down\up}^\mu\bar{\rho}_{\up})^2
  }{
    \langle\langle 
    \calJ_\mu
    \rangle\rangle\langle\langle
    \calJ_{\nu}
    \rangle\rangle
  }
  \,,
\end{align}
where we noted $\bar{\rho}_\up+\bar{\rho}_\down=1$. In terms of the conditional density matrices
Eq.~\eqref{eq:RhoAfterScatterEvent}, we have
\begin{align}
  g^{(2)}_{\mu\mu}(0) - 1
  &=
  \frac{
    -\langle\langle \calJ_\mu \rangle\rangle_{c,\mu} + \langle\langle \calJ_\mu \rangle\rangle
  }{
    -\langle\langle \calJ_\mu \rangle\rangle
  }
  \\
  &=
  \frac{
    \Gamma^\mu_{\downarrow\uparrow}(\bar{\rho}^{c,\mu}_\uparrow - \bar{\rho}_\uparrow)
    +
    \Gamma^\mu_{\uparrow\downarrow}(\bar{\rho}^{c,\mu}_\downarrow - \bar{\rho}_\downarrow)
  }{
    -\langle\langle \mathcal{J}_\mu \rangle\rangle
  }
  \,.
\end{align}
Here, $\langle\langle \calJ_\mu \rangle\rangle_{c,\mu}=\Tr[J_\mu{}\rho^{c,\mu}]=\langle\langle{J_\mu{}J_\mu}\rangle\rangle/\langle\langle{J_\mu}\rangle\rangle$.
The denominators above are all positive.  Note that elastic processes
cancel here, because they do not change the state of the QD and so
cannot contribute correlated events. The bunching behavior is thus
obtained by comparing rates of two consequent events to the
steady-state rates of the single events, or equivalently, whether the
conditional total rate of inelastic electron-transferring scattering
events increases or decreases.  Finally, consider the situation where
one of the two rates is larger than the other (so that
e.g. $(\Gamma_{\uparrow\downarrow}^\mu)^2>\Gamma_{\uparrow\downarrow}^\mu\Gamma_{\downarrow\uparrow}^\mu>(\Gamma_{\downarrow\uparrow}^\mu)^2$).
If the steady-state spin polarization does not significantly block the
fast process (e.g. $\bar{\rho}_\downarrow\sim{}1$),
$g^{(2)}_{\mu\mu}(0)<1$ from Eq.~\eqref{eq:IntG2MinusOneByRates} and
we find sub-Poissonian noise. In contrast, if the spin polarization
blocks the fast process ($\bar{\rho}_\downarrow\sim{}0$), then
$g^{(2)}_{\mu\mu}(0)>1$ and the noise becomes super-Poissonian.  In
summary, we find that the sub/super-Poissonian noise regimes arise due
to a combination of the relative magnitudes of the rates of the QD
relaxation/excitation processes, and the polarization of the QD
blocking some of them, which depends on the sign of the bias voltage.

\subsection{Effects of electron-electron interactions in the edge channel}

We can also include electron-electron interaction in the helical edge
state (the case $K<1$), which modifies the tunneling rates, which are proportional to
the lead correlation functions $F$.  For $|\omega-\sigma_\alpha
eV|>\kbT$, the correlation functions behave as
\begin{multline}
  \label{eq:CorrFuncPowerLaw}
  F_{\alpha\bar\alpha}(\omega)
  \propto
  \big((\omega-\sigma_\alpha eV)\beta/2\pi\big)^{2K-1}
  \\ \times
  \begin{cases}
    1 & \omega-\sigma_\alpha eV>0 \\
    \exp\big((\omega-\sigma_\alpha eV)\beta\big) & \omega-\sigma_\alpha eV<0
  \end{cases}.
\end{multline}
Without interactions, this function transitions from exponential to
linear behavior. With interactions, the linear behavior is replaced by
an algebraic behavior with an exponent of $2K-1$, and for $K<1$ this generally
shifts the weight of the rates towards $\omega\approx\sigma_\alpha eV$.

In Fig.~\ref{fig:FanoDecompositionWithInteraction}, the Fano factor
and the different terms from Eq.~\eqref{eq:FanoByG2Function} are
shown, including interaction in the edge states. The
region of the sub-Poissonian noise at $V>0$ is reduced. By looking at the
contribution of the correlated scattering events (red dashed), we
however, find that this occurs even though the events are antibunched.
A closer look at the weight factors $|I_\pm/\bar I|$ shows that at $eV\approx\Delta_Z$, there is a total increase in the rate of
scattering events that reduce the number of left-movers ($|I_-/\bar{I}|$), even
though the voltage is positive, such that the ratio
$(I_+-I_-)/\bar{I}=(|I_+|+|I_-|)/\bar{I}$ increases. Similarly as in
the thermal noise region, this results to $F>1$ even though the
events are uncorrelated.  As the sub-Poissonian behavior is
compressed towards $eV\approx\Delta_Z$ due to the power law behavior
of the transition rates, the antibunching is almost not visible in the
Fano factor. Hence, the system here is an example of the problem mentioned in Sec.~\ref{sec:jump-operators}:
in the presence of multiple types of scattering events, the value of the Fano factor may be unrelated to information
about the correlations between the events, even in a parameter region nominally away from the thermal noise region.

This behavior originates from the process described by
$\Gamma_{\down\up}^-\rho_\up$, which is a relaxation process that
decreases the number of left movers. It is a spin conserving relaxation process,
so its rate is large and it is energetically allowed for $eV<\Delta_Z$. For $|eV|<\Delta_Z$, the
QD polarization approaching the ground state however suppresses its contribution to $I_-$.
This suppression is weakest for $eV\approx\Delta_Z$, where the rate moreover is in a transition region
between algebraic and exponential behavior, which results to the nonmonotonicity in $I_-$.
This transition region has a width of $\kbT$, which determines the magnitude of the feature.
For the polarization and the other rates, however, the behavior is
determined by $|eV|/\Delta_Z$ such that the region of the sub-Poissonian noise
increases for increasing $\Delta_Z$. By choosing the Zeeman splitting large
enough the sub-Poissonian behavior can thus be restored for the interacting problem.

\section{Conclusion}
We have studied the noise properties of a spin 1/2 impurity, e.g. in a quantum dot, weakly coupled to a helical edge of a two-dimensional topological insulator. In a setup where the spin is subject to a Zeeman field but where the effect of such a field is negligible in the helical edge, the Fano factor of backscattering events for electrons in the helical edge states shows a characteristic antibunching ($F<1$) to bunching ($F>1$) transition behavior with sweeping the bias voltage in the helical edge in a regime where the bias voltage is comparable to the Zeeman splitting. We show that this transition of the noise properties is only present when the tilt-angle $\theta_Z$ of the magnetic field with respect to the quantization axis of the helical edge states satisfies $0<\theta_Z<\pi/2$. Note that at $\theta_Z=\pi/2$, the specifics of spin helical leads is effectively absent in the backscattering current, since the spin exchange with the spin 1/2 impurity is not locked to the direction of momentum exchange. The antibunching to bunching transition in the Fano factor is therefore a unique signature of helical edge states. 

\acknowledgments

P.V. acknowledges funding from EU's Horizon 2020 research and
innovation program under Grant Agreement No. 800923 (SUPERTED). P.R. acknowledges financial support by the Deutsche Forschungsgemeinschaft (DFG, German Research Foundation) within the framework of Germany’s Excellence Strategy–EXC-2123 QuantumFrontiers–390837967.

\end{document}